# *Improving the detection accuracy of unknown malware by partitioning the executables in groups*

Ashu Sharma [1,1], Sanjay K. Sahay [1] and Abhishek Kumar [1],

[1] Department of Computer Science and Information System,
BITS PILANI, K. K. Birla Goa Campus
Zuarinagar-403726, Goa, India
{p2012011,ssahay,f2010490}@goa.bits-pilani.ac.in

**Abstract.** Detection of unknown malware with high accuracy is always a challenging task. Therefore, in this paper we study the classification of unknown malware by two methods. In the first/regular method, similar to other authors [17][16][20] approaches we select the features by taking all dataset in one group and in second method, we select the features by partitioning the dataset in the range of file 5 KB size. We find that the second method detect the malwares with ~ 8.7% more accurate than the first/regular method.

## 1 Introduction

Malwares are continuously evolving and are big threats to the leading Windows and Android computing platforms [3]. The attacks/threats are not only limited to individual level, but there are state sponsored highly skilled hackers developing customized malwares [25], [7], [10]. These malwares are generally classified as a first and second-generation malwares. In first generation, structure of the malwares remains constant, while in second generation, structure changes in every new variant, keeping the action same [12]. On the basis of how variants are generated in malware, second generation malwares are further classified into Encrypted, Oligormorphic, Polymorphic and Metamorphic Malwares [24].

Its an indisputable fact that the prolong traditional approach (signature matching) of combating the threats/attack with a technology-centric are ineffective to detect second generation customized malwares. If in time adequate measures has not taken, the consequence of the scale (more 317 million new malwares are reported [1] in the year 2014) at which malware are developed will be very devastating. Nevertheless, the second-generation malware are very effective and not easy to detect. Recently a new malware is reported by McFee which is capable to infect the hard drives and solid state storage device (SSD) firmware and the infection cannot be removed even by formatting the devices or reinstallation of operating systems [2]. Therefore, there is a need that both academia and anti-malwares developers should continually work to combat the threats/attacks from the evolving malwares. The most popular techniques used for the detection of malwares are signature matching, heuristics based detection, malware normalization, machine learning, etc [24].

In recent years, machine learning techniques are studied by many authors and proposed different approaches [4] [11] [19], which can supplement traditional anti-malware system (signature matching).

---

[1] p2012011@goa.bits-pilani.ac.in

Hence, in this paper for detection of unknown malware with high accuracy, we present a static malware analysis method which detect the unknown malware with ~ 8.7% more accurate then the regular method. The paper is organized as follow, in next section related work is discussed. In section 3 we discuss the data preprocessing and feature selection. Section 4 contains the brief description of Naive Bayes classifier. In section 5 we discuss the method to improve the detection accuracy and results. Finally section 6 contains the conclusion and future direction of the paper.

# 2 Related Works

To combat the threats/attacks from the second generation malwares, Schultz et al. (2001) was the first to introduce the concept of data mining to classify the malwares [23]. In 2005 Karim et al. [14] addressed the tracking of malware evolution based on opcode sequences and permutations. In 2006, O. Henchiri et al. [13] reported 37.17% detection accuracy from a hierarchical feature extraction algorithm by using NB (Naive Bayes) classifier. Bilar (2007) uses small dataset to examine the opcode frequency distribution difference between malicious and benign code [8] and observed that some opcodes seen to be a stronger predictor of frequency variation. In 2008, Yanfang Ye et. al. [29] applied association rules on API execution sequences and reported an accuracy of 83.86% with NB classifier. In 2008, Tian et al. [27] classified the Trojan using function length frequency and shown that the function length along with its frequency is significant in identifying malware family and can be combined with other features for fast and scalable malware classification. Moskovitch et al. (2008) studied many classifier viz. NB, BNB, SVM, BDT, DT and ANN by byte-sequence n-grams (3, 4, 5 or 6) and find that NB classifier detect the malwares with 84.53% accuracy [17]. In 2009 S. Momina Tabish [26] used 13 different statical features computed on 1, 2, 3 and 4 gram by analysing byte-level file content for classification of malwares. In 2009 Syed Bilal Mehdi et al. [16] obtained 64% and 58% accuracy by NB classifier with good evaluator feature selection scheme on 4 gram and 6 gram features from the executables. Chandrasekar Ravi et al. in year 2012 reported 48.69% accuracy with NB classifier by using 4-grams Windows API calls features [20]. Chatchai Liangboonprakong et al. (2013) proposed a classification of malware families based on N-grams sequential pattern features [15]. They used DT, ANN and SVM classifier and obtained good accuracy. In 2013 Santos et al. [22] used Term Frequency for modelling different classifiers and among the studied classifier, SVM outperform for one opcode and two opcode sequence length respectively. Recently (2014) Zahra Salehi et al. construct feature set by extracting API calls used in the executables for the classification of malwares [21].

# 3 Data Preprocessing and Feature Selection

For the experimental analysis, we downloaded 11088 malwares from malacia-project [18] and collected 4006 benign programs (also verified from virustotal.com [9]) from different systems. In the collected dataset we found that 97.18% malwares are below 500 KB, (fig. 1) hence for the classification we took the data set which are below 500 KB.

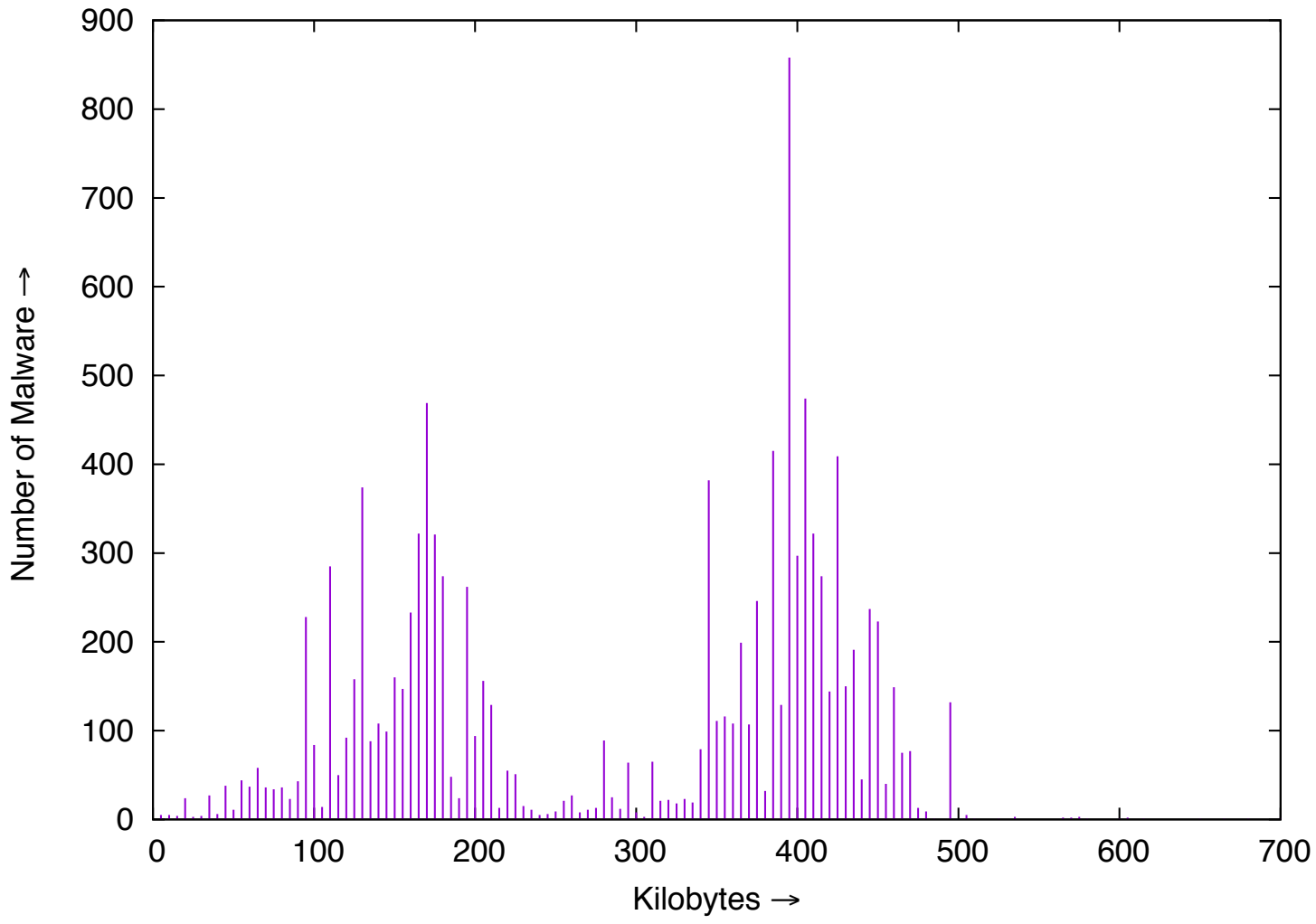

Fig. 1: *Number of malwares in the group of 5 KB size.*

For classification the features are opcodes of executables obtained by objdump utility available in the linux system and its selection procedure is given in the algo. 1. To test our methods we select popular Naive Bayes classifier which can handle an arbitrary number of independent variables and is briefly described in next section.

---

**Algorithm 1: Feature Selection**

---

**INPUT:** Pre-processed data
$\mathbf{N_b}$: Number of benign executables, $\mathbf{N_m}$: Number of malware executables, **n**: Number of features required
**OUTPUT:** List of features
**BEGIN**
**for all** benign data **do**
Add all frequency $\mathbf{f_i}$ of each opcode **o** and Normalize them with respect to $\mathbf{N_b}$

$$F_b(o_j) = (\sum f_i(o_j)) / N_b$$

**end for**
**for all** malware data **do**
Add all frequency $\mathbf{f_i}$ of each opcode **o** and Normalize them with respect to $\mathbf{N_m}$

$$F_m(o_j) = (\sum f_i(o_j)) / N_m$$

**end for**
**for all** opcode $\mathbf{o_j}$ **do**
Find the difference of each opcode normalized frequency $\mathbf{D(o_j)}$.

$$D(o_j) = \left| F_b(o_j) - F_m(o_j) \right|$$

**end for**
**return** **n** number of opcodes with highest **D(o)**.

---

# 4 Naive Bayes classifier

Given a set of features (opcodes), $O = o_1, o_2, o_3 ..., o_n$, the Naive Bayes classifier gives the posterior probability for the class C (malware/benign) and can be written as:

$$P(C \mid o_1, o_2, ...o_n) = \frac{P(C)P(o_1, o_2, ...o_n \mid C)}{P(o_1, o_2, ...o_n)} \tag{1}$$

where $P(C|o_1,o_2,..o_n)$ is posterior probability of the class membership, i.e., the probability that a test executable belongs to class C, Since Naive Bayes assumes that the conditional probabilities of independent variables are statistically independent, we can decompose the likelihood to a product of terms:

$$P(o_1, o_2, ...o_n \mid C) = \prod_{i=1}^{n} P(o_i \mid C) \tag{2}$$

Here, $P(o_i|C)$ is the probable similarity of occurrence of feature $o_i$ of the class $C$ and can be computed by the equation:

$$P(o_i \mid C) = \frac{1}{\sqrt{2\pi\sigma_C^2}} e^{-\frac{(o-\mu_C)^2}{2\sigma_C^2}} \tag{3}$$

where $o$ denotes the feature $o_i$ opcode of the test executable and $\mu_c$, $\sigma_c$ are the mean and variance of the class $C$.

From the above, the final classification can be done by comparing the posterior probability between both class models, if malware class posterior probability of test executable is high then it is classified to malware else classified as benign.

# 5 Method to Improve the Detection Accuracy of Unknown Malware

We first study the regular method and then propose our novel partitioned method i.e. grouping the executables, to improve the detection accuracy of unknown malware. Both method are depicted in the fig. 4 and 6.

## 5.1 Regular method

Fig. 4 represents the regular method for finding the features for the classification of unknown malwares. In this method, for the classification, we train the classifier by 10558 malwares and 2454 benign executables. For the purpose, the procedure to obtain the features of the dataset is given in algo. 1.

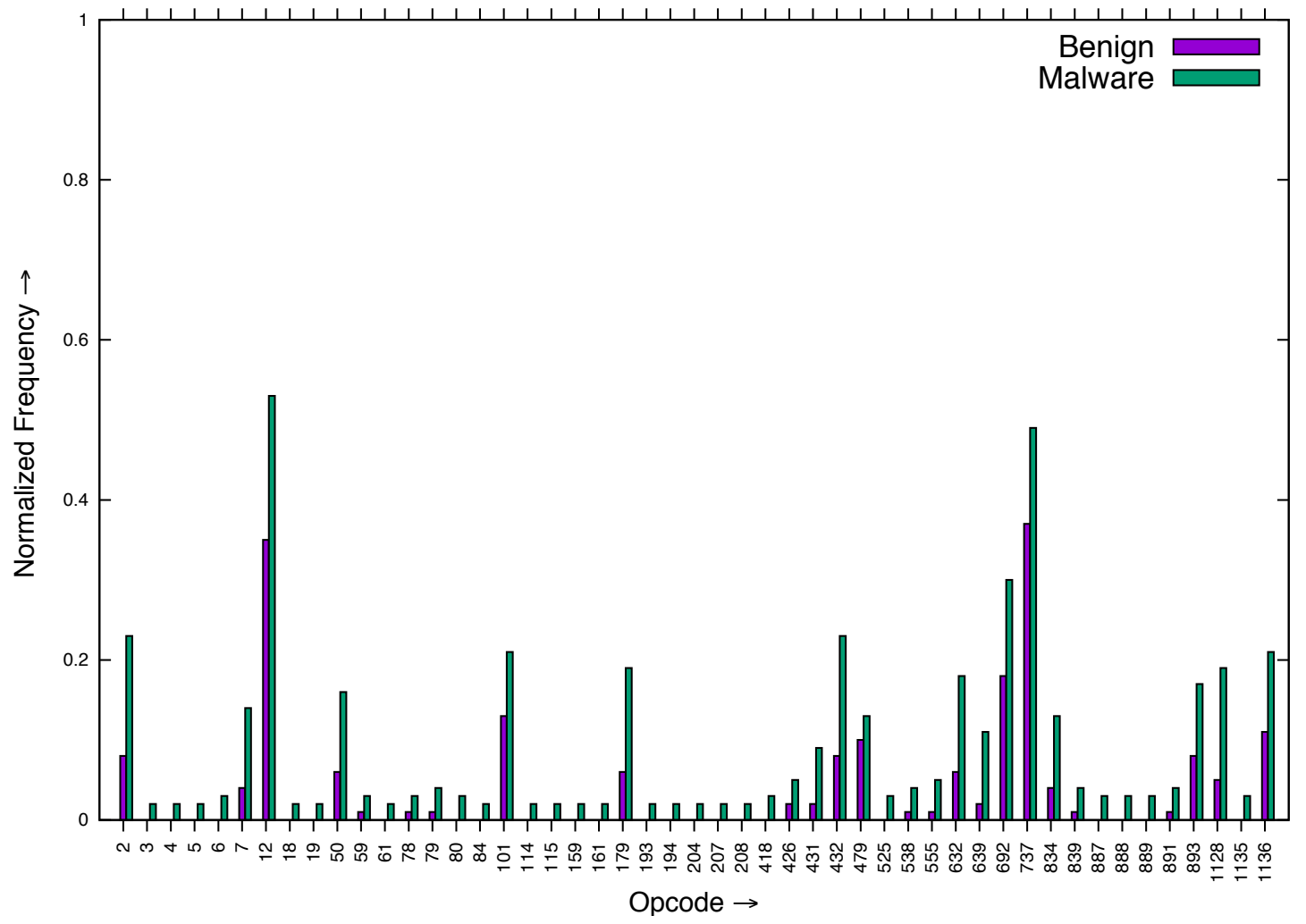

Fig. 2: *Opcodes which found more in malwares.*

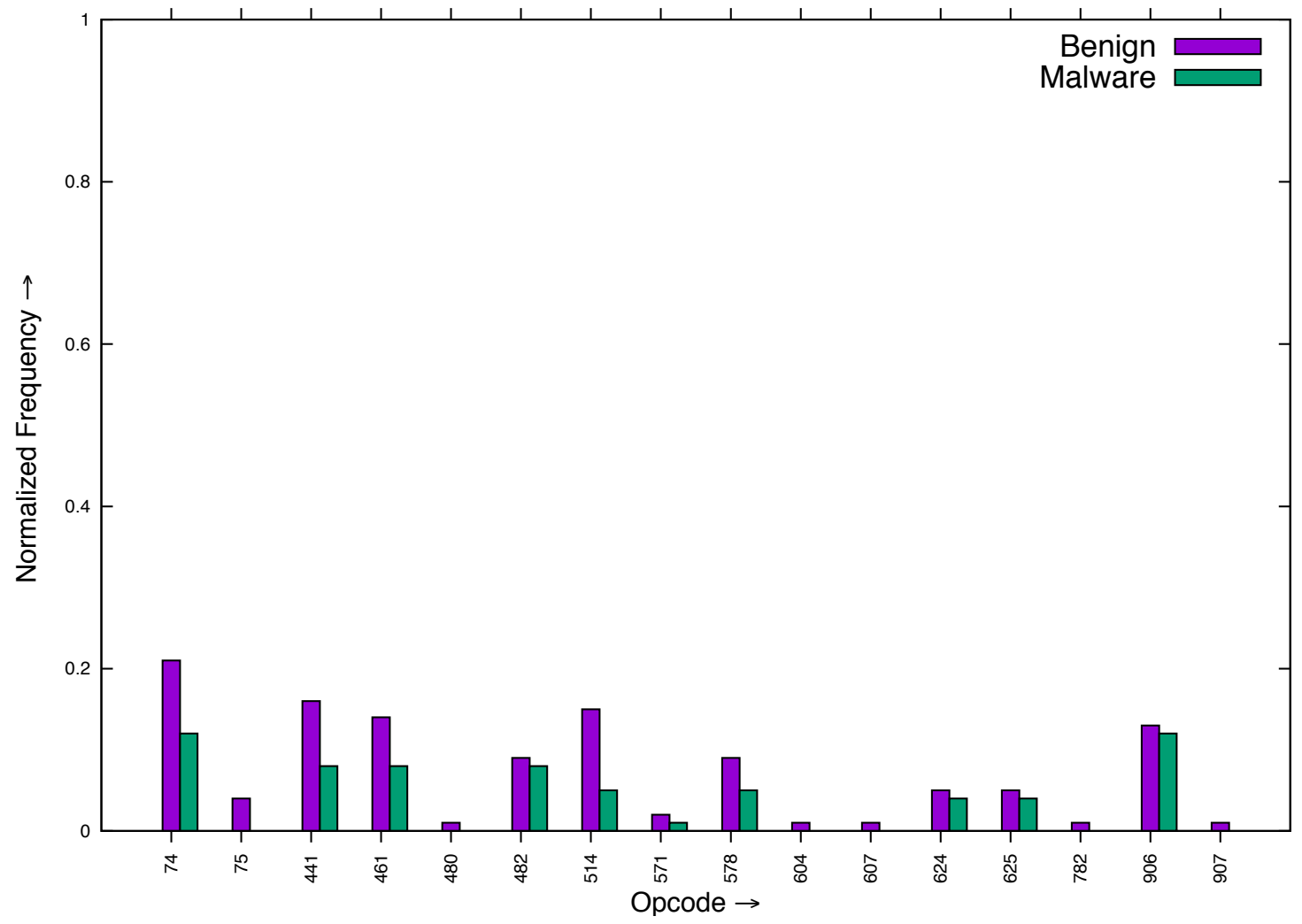

Fig. 3: *Opcodes which found more in benign programs*

We study opcode occurrence of the collected dataset and found (fig. 2 and 3) that malware opcodes distribution and/or occurrence differ significantly from benign program. Hence, we obtained the promising features by computing the difference of normalized opcodes frequency between benign and malware executables.

We use the machine learning tool WEKA to train and test the chosen (NB) classifier. For the testing 750 malware 610 benign executables are taken from the dataset which are not used for training the classifier. To evaluate classifiers capability, we measure detection accuracy, i.e., the total number of the classifier's hits divided by the number of executables in the whole dataset and is given as:

$$Accuracy(\%) = \frac{TP + TN}{TM + TB} \times 100 \quad (4)$$

where,
$TP \longrightarrow$ True positive, the number of malwares correctly classified.
$TN \longrightarrow$ True negative, the number of benign correctly classified.
$TM \longrightarrow$ Total number of malwares.
$TB \longrightarrow$ Total number of benign.
The accuracy obtained by this method by increasing the features in systematic way are shown in fig. 7. We find that the accuracy of the classifier is almost flat, if the number of features are greater than 90. The best accuracy obtained by this method is 78.33%.

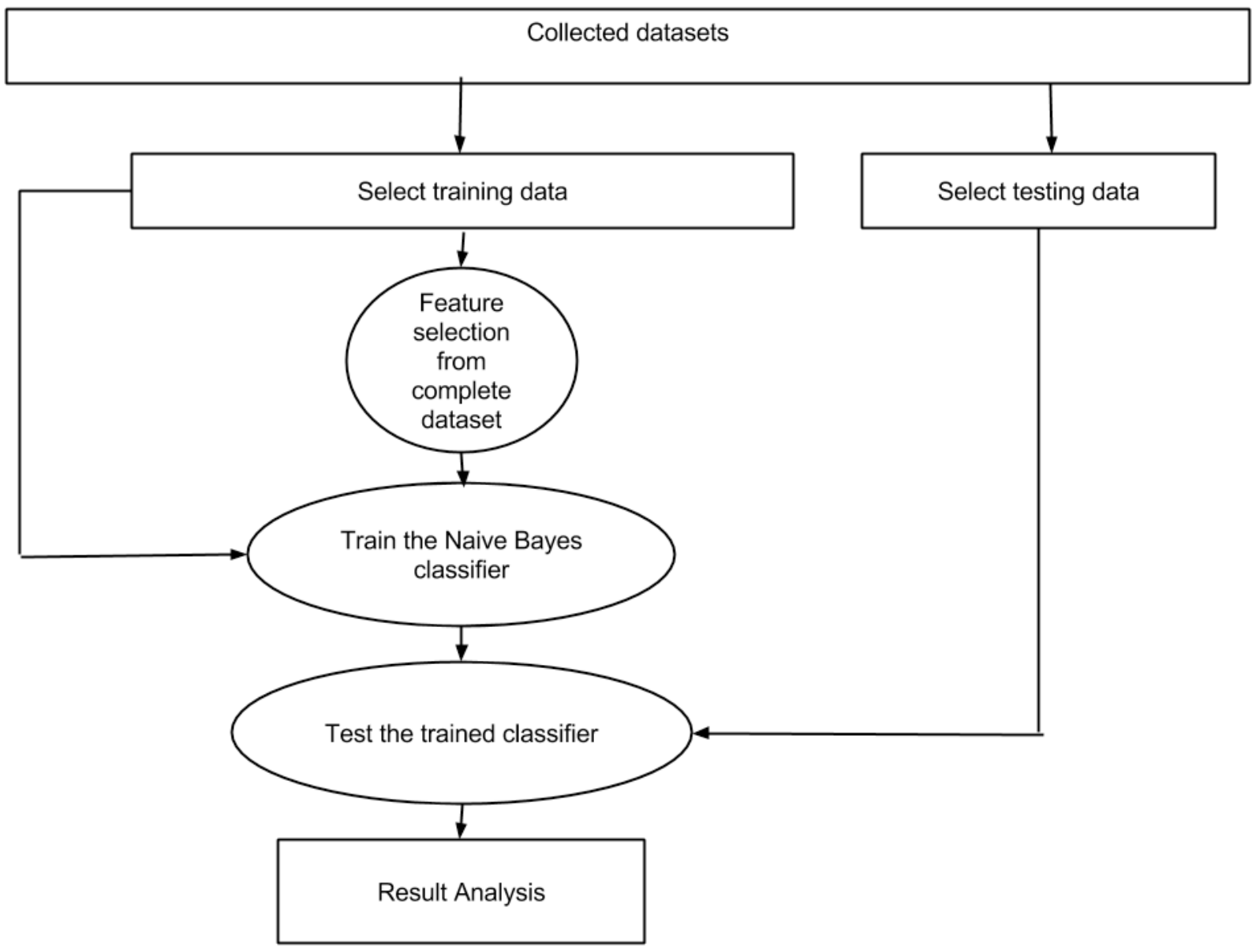

Fig. 4: *Flow chart for the detection of unknown malwares without partitioning method.*

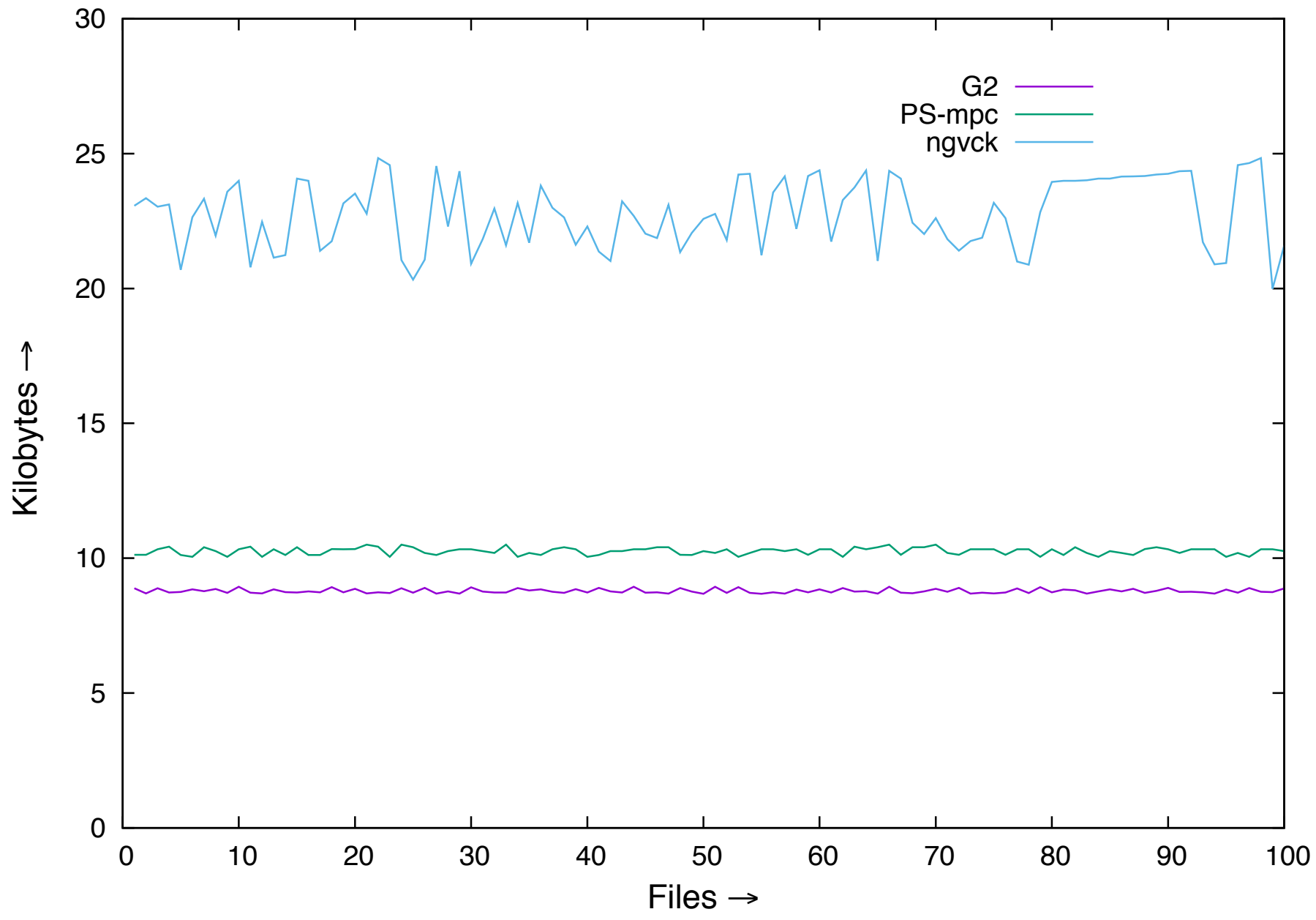

Fig. 5: *Fluctuations in the size of malwares generated by G2, PS-MPC and NGVCK.*

## 5.2 Partitioning method

In this method as shown in fig. 6, we partitioned the collected dataset in 5KB size range. The partition size is based on the study that size of any two malwares generated by G2 [5], PS-MPC [6] and NGVCK [28] kits does not vary by more than 5 KB size (fig. 5).

For the comparative experimental analysis we took same classifier, training and testing data which are used in the regular method. In this method procedure of feature selection is same as given in algo. 1. However, to improve the accuracy of detection we first listed the features of each group given by algo. 1, then serially trained and tested the chosen classifier by systematically increasing the features. We found that the accuracy obtained by this method outperformed the regular method (fig. 7) and the best accuracy obtained is 87.02% i.e. the detection accuracy is ~ 8.7% more compared to regular method.

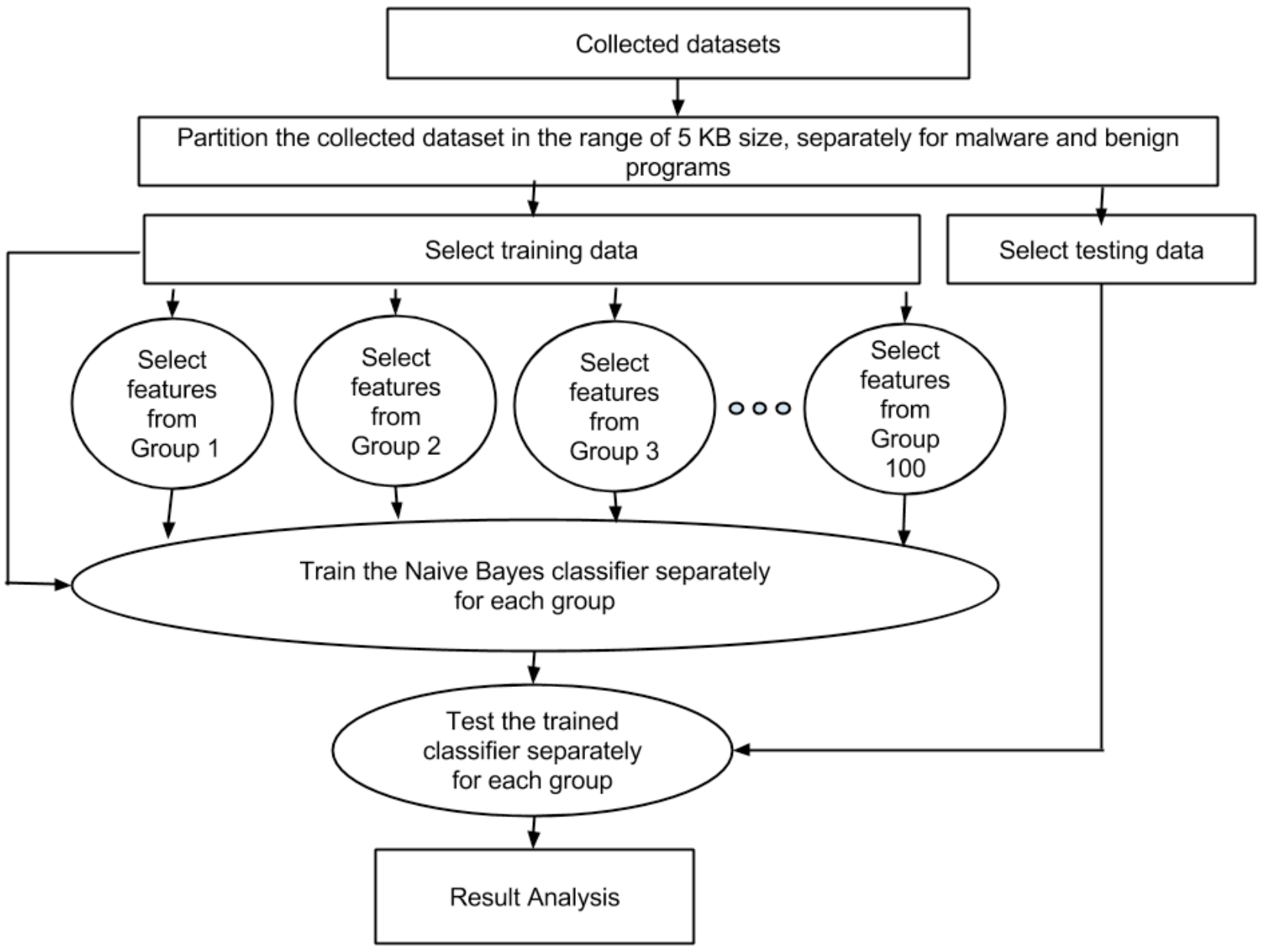

Fig. 6: *Flow chart for the detection of unknown malwares by partitioning the executables in groups.*

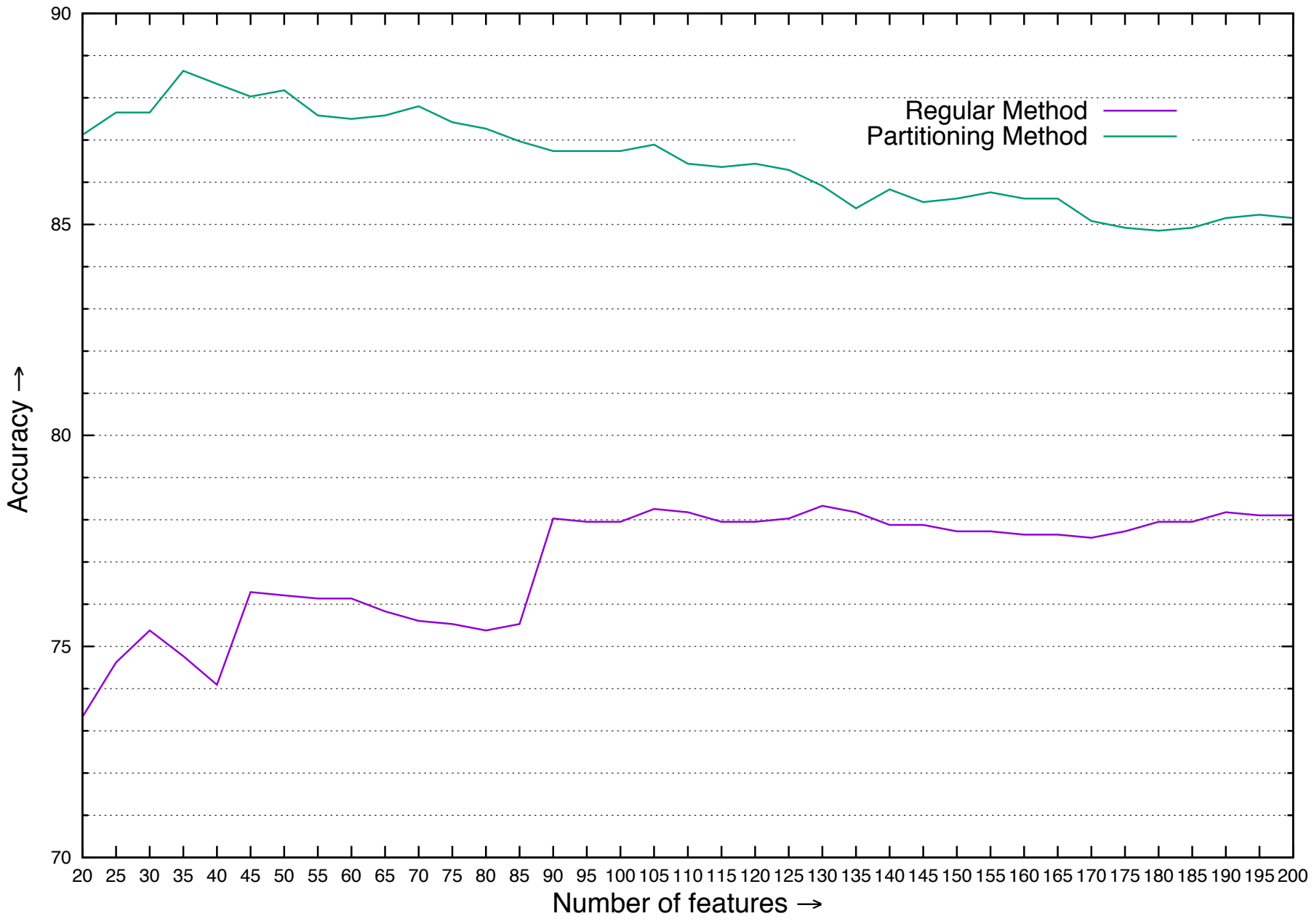

Fig. 7: *Detection accuracy obtained by both the methods.*

## 6 Conclusion

In this paper we present a novel method to improve the detection accuracy of unknown malwares. We first investigated the variation in the size of malwares generated by G2, PS-MPC and NGVCK and found that the generated malware size vary from 0 to 5 KB. Therefore we partitioned the collected dataset in 100 groups, each in 5 KB range of size and then selected the feature from each group to train classifier for the detection of unknown malwares. We found that, if features are selected by partitioning dataset in the range of 5 KB then the malwares are detected with ~ 8.7% more accurate then the regular method. As the result is very promising, in future, we will study the partitioning method in-depth for the further improvement in the detection accuracy of unknown malwares.

## Acknowledgment

Mr. Ashu Sharma is thankful to BITS, Pilani, K.K. Birla Goa Campus for the support to carry out his work through Ph.D. scholarship No. Ph603226/Jul. 2012/01. We are also thankful to IUCAA, Pune for providing hospitality and computation facility where part of the work was carried out.